\documentclass[twocolumn,samsmath,amssymb,floatfix,superscriptaddress,aps]{revtex4}
\usepackage{graphicx}
\begin{document}
\title{From compact to fractal crystalline clusters in concentrated systems of monodisperse hard spheres}

\author{Chantal Valeriani} \affiliation{Departamento de Quimica Fisica, facultad de Ciencias Quimicas, Universidad Complutense, 28040 Madrid, Spain}
 \affiliation{SUPA, School of Physics and Astronomy, University of Edinburgh, Mayfield Road, Edinburgh, EH9 3JZ, Scotland}
\author{Eduardo Sanz} \affiliation{Departamento de Quimica Fisica, facultad de Ciencias Quimicas, Universidad Complutense, 28040 Madrid, Spain}
 \affiliation{SUPA, School of Physics and Astronomy, University of Edinburgh, Mayfield Road, Edinburgh, EH9 3JZ, Scotland}
\author{Peter N. Pusey} \affiliation{SUPA, School of Physics and Astronomy, University of Edinburgh, Mayfield Road, Edinburgh, EH9 3JZ, Scotland}
\author{Wilson C. K. Poon} \affiliation{SUPA, School of Physics and Astronomy, University of Edinburgh, Mayfield Road, Edinburgh, EH9 3JZ, Scotland}
\author{Michael E. Cates} \affiliation{SUPA, School of Physics and Astronomy, University of Edinburgh, Mayfield Road, Edinburgh, EH9 3JZ, Scotland} 
\author{Emanuela Zaccarelli} \affiliation{ {CNR-ISC and Dipartimento di Fisica, 
Universit\`a di Roma La Sapienza, P.le A. Moro
  2, 00185 Roma, Italy} }

\begin{abstract}
We address the crystallization of monodisperse hard spheres in terms of the properties of finite-size crystalline clusters. By means of large scale event-driven Molecular Dynamics simulations, we study systems at different packing fractions $\phi$ ranging from weakly supersaturated state points to glassy ones, covering different nucleation regimes. We find that such regimes also result in different properties of the crystalline clusters:
compact clusters are formed in the classical-nucleation-theory regime ($\phi \le 0.54$), while a crossover to fractal, ramified clusters is encountered upon increasing packing fraction ($\phi \ge 0.56$), where nucleation is more spinodal-like. We draw an analogy between macroscopic crystallization of our clusters and percolation of attractive systems to provide ideas on how the packing fraction influences the final structure of the macroscopic crystals.
In our previous work~({\em Phys.~Rev.~Lett.}, {\bf 106}, 215701, 2011), we have demonstrated how crystallization
from a glass (at $\phi>0.58$) happens via a gradual (many-step) mechanism: in this paper we show how the 
 mechanism of gradual growth seems to hold also in super-saturated systems just above freezing showing that static properties of  
clusters are not much affected by dynamics.

\end{abstract}

\maketitle

\section{Introduction \label{sec:intro}}

Following the pioneering computer simulations of Alder and Wainwright~\cite{alder}, more than 50 years ago, assemblies of hard spheres in thermal motion have become the subject of intense research aimed at unveiling the fundamental physics of both thermodynamic and kinetic phase transitions.  This numerical work has been complemented by experimental studies of suspensions of ``hard-sphere'' colloidal particles (see Ref.~\cite{peterbill,puseyreview,vincent,weeksjpcm}).



Despite the simplicity of the hard-core interactions, the behaviour of 
this system is far from trivial and several important questions still remain unanswered. Concerning thermodynamics, the location of the fluid-to-crystal transition, occurring in the  packing
 fraction $\phi$ range $0.492 < \phi < 0.543$  has been quite well characterised~\cite{alder,noyavega,verrocchio,dullens}. 
However, open issues remain concerning the effect of polydispersity~\cite{sollich,wildingnew,ito}
 and, most importantly, the reported discrepancy between 
calculated nucleation rates and those measured experimentally~\cite{auer}. These issues have recently caused a  
revival of studies on hard-sphere nucleation~\cite{PuseyRS,filion1,filion2,schillinglast}. In particular, 
recent works aim to elucidate the mechanism of crystal formation in terms of preferred precursor 
structures~\cite{tanaka,tanakanatmat,tanakajpcm}, proposing a two-step scenario~\cite{schope,schilling}, 
and also aim to assess the competition between different solid polymorphs~\cite{john}.

Not only does the fluid-to-crystal transition remain debated, 
but also the presence of an ideal glass transition at packing fractions 
below random close packing ($\phi_{rcp} \approx 0.64$) is still under discussion. 
The initial observations by Pusey and van Megen~\cite{peterbill} 
of a glass transition taking place for $\phi \sim 0.58$ were later 
challenged by works in microgravity~\cite{kegel,chaikin} and by 
numerical simulations by Torquato and coworkers~\cite{torquato}. 
Moreover, a recent work by  Brambilla et al.~\cite{brambilla} caused considerable 
debate in the glass community~\cite{commentvanmegen,commentfuchs,medinanoyolapre}. 
By considerably extending the observation time window, 
these authors suggested that the glass transition was always obviated by activated 
processes whenever $\phi < \phi_{rcp}$. However,  delicate issues remain, concerning for example the 
correct way of estimating the ``true'' colloidal packing 
fraction in these experiments~\cite{poonweeksroyallpreprint}. 

With this as background, we have recently started an extensive numerical investigation 
of hard-sphere systems, via event-driven Molecular Dynamics simulations. 
Initially, we assessed the role of polydispersity on dynamics and crystallization~\cite{ZaccarelliPRL,PuseyRS},  finding that, while changing substantially the nucleation behaviour, 
a small degree of polydispersity $s\lesssim 6\%$ does not strongly affect the dynamics, 
leaving unchanged the tendency to form an ideal glass around 0.58~\cite{ZaccarelliPRL}. 
In addition, we identified two regimes of nucleation~\cite{PuseyRS,chantaljpcm}: 
standard nucleation and growth at concentrations in and slightly 
above the coexistence region and, beyond $\phi \sim 0.56$, ``spinodal nucleation'', 
where the free energy barrier to nucleation appears to be negligible. In the latter regime, at very high volume fractions, we have investigated the subtle 
interplay between slow dynamics and crystallization, elucidating a novel mechanism by which hard-sphere glasses crystallize without 
particle diffusion (even at the single particle level) on scales comparable to or beyond their radius~\cite{sanzPRL}.
Diffusionless crystallisation has also been recently reported for a 3 dimensional lattice system~\cite{eli}.

In this work, we investigate an aspect of crystallisation of monodisperse hard spheres 
that, to our knowledge, has not been studied so far: the statistics properties of crystalline clusters as 
crystallisation proceeds. 
By monitoring a large system, we identify  the solid-like clusters 
and analyse their size distribution and shape 
as a function 
of packing fraction and of time.
These clusters undergo a percolation transition of the kind described 
in simulation~\cite{morris} and in theoretical work on molecular 
systems~\cite{wolynes}.
We monitor the cluster size distribution and clusters radius of gyration as a function of time, 
to address whether non-trivial exponents can be extracted and related to the standard 
percolation scenario that is commonly encountered in attractive systems~\cite{stauff}.

A comment is due on the definition of clusters of hard-sphere particles, 
where no attraction is present.
Indeed, clusters are rigorously defined~\cite{satorreview} for attractive 
systems  in terms of an energy scale which cohesively keep 
particles together. Moreover, in the case of colloidal particles 
interacting with competing short-ranged attraction (e.g. depletion) 
and long-ranged repulsion (induced by electrostatics), clusters are found 
to be stable minima of the underlying potential~\cite{groenewold,sciortinoPRL2004,mossa,archerwilding}. 
These equilibrium clusters have been observed
in both experiments and simulations~\cite{stradner,campbell,zaccaclu,cardinauxEPL} and 
their role as a building blocks of non-equilibrium structures like Wigner glasses and gels is still under 
active investigation~\cite{sciortinoPRL2004,zaccareview,charbonneaupre,toledano,klixroyalltanaka,cardinauxjpcb,yukawaescv}.
In the case of hard spheres, particles forming a cluster "glue" due to the thermodynamic 
driving force toward crystal nucleation~\cite{kelton,debenedetti} rather than due to attractive interactions between particles.
Hence, whenever we find two nearest-neighbour 
solid-like particles (according to the definitions discussed below), we consider them 
as belonging to the same crystalline cluster. Of course, these clusters 
can be either transient or permanent, in analogy with attractive clusters at finite temperature. 

Within our study, we cannot monitor the equilibrium distribution of such clusters 
because of the irreversible process of crystallization once it is triggered. However, we can 
follow their growth while crystallization  proceeds, so that with time clusters become 
larger and larger and consequently more and more permanent, until macroscopic crystallization 
of the samples occur. In particular, we study crystallization at
different packing fractions, in order to address the question of whether significant 
differences can be found between the different nucleation regimes. Our findings 
suggest an important crossover from  compact to fractal clusters with increasing 
supersaturation, which reflects the change in the underlying crystallisation processes. 
These findings could have important consequences for the final structures of the resulting macroscopic crystals.

The manuscript is organised as follows: we first present the simulation and analysis method in Section~\ref{simu}; next, 
we report our results on  the clusters structure at different packing fractions 
(Section~\ref{sec:shape}), the percolation of the crystalline clusters 
(Section~\ref{sec:percolation}),  the cluster size distribution (Section~\ref{sec:csd})
and on the effect of the criterion 
to identify solid-like particle on properties of the growing crystalline clusters
(Section~\ref{sec:nsc}).

\section{Simulation and Analysis Methods\label{simu}}

We perform event-driven Molecular Dynamics simulations in the $NVT$ ensemble with cubic periodic
 boundary conditions for a large system of N = 86400 monodisperse hard spheres~\cite{rapaport,zacca1}. 
Mass, length, and time are measured in units of particle mass $m$, particle diameter $\sigma$ and 
$\sqrt{m \sigma^2/\kappa_B T}$, where $\kappa_B$ is the Boltzmann constant and $T$ the temperature 
and we set $\kappa_B T=1$.  
The system is prepared at different packing fractions $\phi=\frac{\pi}{6} N \sigma^3/V$ (with V the system's volume) beyond freezing ranging from $\phi=0.54$ to  $\phi=0.61$. 
At $\phi \ge 0.54$, the metastable fluid-phase spontaneously crystallizes within the duration of our simulations. With increasing $\phi$, the relaxation of the system becomes slower and slower, until the glass transition around $\phi \sim 0.58$ is encountered~\cite{ZaccarelliPRL}. Above this value, ageing effects are present and crystallization occurs from a glass~\cite{sanzPRL}.

To study the formation and growth of crystalline clusters, 
we first identify the solid particles in the system. 
To this end, we use the rotationally invariant local bond order parameter 
$d_6$ defined as the scalar product between particle $i$'s $q_6$ 
complex vector and the one of each of its neighbour $j$, 
$d_6(i,j)$ (see Appendix~\ref{sec:appa} for further details).
 Once we have identified all solid particles $N_s$ in the system, 
we monitor the fraction of crystalline particles, defined as $X=\frac{N_s}{N}$.
To identify crystalline clusters, we run a cluster-algorithm: starting 
from a solid particle, we locate its solid neighbours within a 
distance of $1.4\sigma$ and define them as belonging to the same cluster.
After having iteratively applied the 
cluster-algorithm to all solid particles in the system, 
at every time-step we identify all crystalline clusters and compute their size ($s$).


To improve the statistics of our results, we consider 10 independent crystallisation trajectories for 
each studied packing fraction,  each initiated from 
configurations without pre-existing crystal nuclei. 
In order to achieve this non-trivial condition, we compress a small system of 400 particles to a high packing fraction, 
$\phi=0.64$, monitoring that the fraction of crystalline particles is less than
 0.005 of the total number of particles. 
We then replicate this system periodically in space, checking that any sign of periodicity resulting from the replication procedure is lost after 
a very short time (as in Ref.~\cite{sanzPRL}). 
Next, before starting Molecular Dynamics runs, we isotropically expand the configuration to the desired packing fraction.

First of all,  we perform an analysis of the structure of the growing crystalline clusters 
at different packing fractions. This is is done both qualitatively, monitoring the structure of the clusters that are visible 
in the snapshots taken from our simulations 
and quantitatively, studying the size-dependence of the cluster's radius of gyration:
\begin{equation}\label{gyration}
R_g = \left[ \frac{1}{s}\sum_{i=1}^{s}  \left|\vec{ r}_i - \vec{R}_{CM} \right|^2 \right]^{1/2}
\end{equation}
where $s$  is the number of particles in the cluster (i.e. the cluster size), 
$\vec{ r}_i$ indicates the position of particle $i$  
 and $\vec{R}_{CM}$ is the position of the centre of mass of the cluster. 
For fractal aggregates, the dependence of the radius of gyration on the cluster size follows a power law, i.e. $R_g \sim s^{1/d_f}$ , where $d_f$ is
the fractal dimension of the aggregation process. For
random percolation in three dimensions, the value of $d_f$ is known to be about 2.5~\cite{stauff}. 
For spherical clusters  $R_g \sim s^{1/3}$, whereas for planar (loose) clusters   $R_g \sim s^{1/2}$ and for linear ones $R_g \sim s$~\cite{stauff,zaccaclu}.

Next, we study the percolation of crystalline clusters by 
estimating the first time  when a system-spanning  cluster appears 
in the simulation box, i.e., the percolation time $\tau_p$. 
In practice, we detect  $\tau_p$ as 
the time when the largest crystalline cluster becomes 
infinite according to our periodic boundary conditions criterion.
After having identified, at each time-step, the largest crystalline cluster,   
we replicate the simulation box and check whether 
we get a single cluster in the replicated system: the time when this happens 
defines $\tau_p$ (which is then averaged over independent runs).
Once $\tau_p$ is known, we estimate the value of the total 
crystallinity  at percolation $X(\tau_p)=X_p$.
Even though we are aware that an accurate location 
of the percolation threshold would require a finite-size scaling 
analysis,  our aim is not to define with high precision the percolation  
threshold, but rather to find a link between the percolation 
transition and macroscopic crystallization.

We also calculate the cluster size distribution, i.e. the fraction of clusters with $s$ solid-like particles. 
The cluster size distribution changes with time during the crystallization process. To reduce its numerical noise,
we evaluate it for configurations of independent trajectories 
having approximately the same crystallinity,
using narrow $X$ intervals of width $\Delta X=0.01$. In this way we can average our results over all trajectories 
within a given $X$-interval and over the 10 independent runs.
According to random percolation theory,  
the cluster size distribution in a three dimensional system 
decays exponentially fast 
for large $s$ below the percolation threshold, whereas it 
follows a power-law behaviour at percolation, i.e.  $s^{-\tau}$ (where $\tau \simeq 2.2$~\cite{stauff}). 


Finally, to  assess the role of the particular choice of parameters 
used to define solid-like clusters, we analyse the effect of the ``solid-like'' criterion on 
the clusters radius of gyration,  
considering different choices of the number 
of solid-like connections $\xi_c$ needed to identify a solid-like particle.     


\section{Results\label{sec:results}}

We present our results as follows: in Section~\ref{sec:shape} 
we perform both a qualitative and a quantitative  analysis of  
the clusters' structure  for packing fractions ranging from  $\phi=0.54$ to $\phi=0.61$; in 
Section~\ref{sec:percolation} we study percolation of the crystalline clusters; 
in Section~\ref{sec:csd} we estimate the cluster size distribution. 
In the last Section,~\ref{sec:nsc}, we analyse the effect of the criterion used 
to identify solid-like particle on properties of the growing crystalline clusters.

\subsection{Structure of the growing crystalline clusters}
\label{sec:shape}

\begin{figure*}[t]
\centering
\includegraphics[width=1\textwidth,clip=]{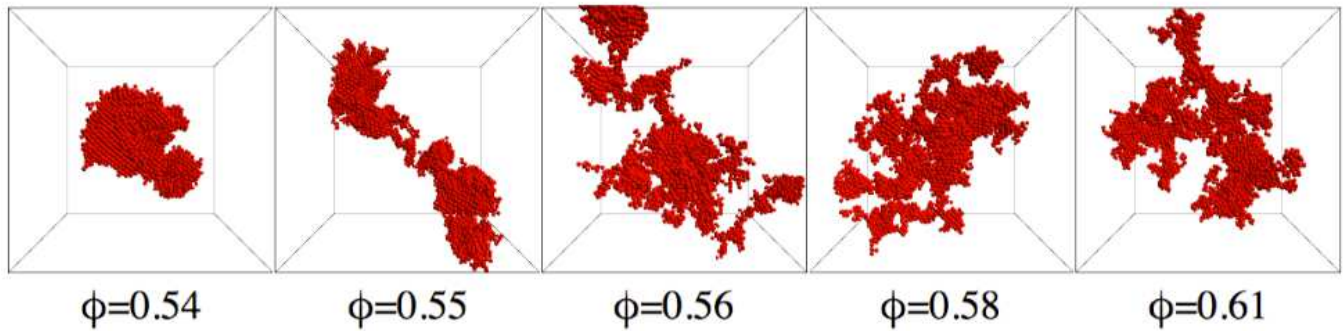}
\caption{Snapshots of typical largest crystalline clusters as a function of $\phi$ at fixed cluster size ($s=5000$).
This size is achieved when $(\phi, X)$ are respectively (0.54, 0.08),  (0.55, 0.15),  (0.56, 0.15), (0.58, 0.14),  (0.61,0.12).
Periodic boundary conditions are taken into account and clusters are centred in the simulation box.}
\label{fig:clusters}
\end{figure*}

First, we study the characteristic structure of growing crystallites at different packing fractions $\phi$.
In order to visualise how the structure of crystalline clusters changes with increasing packing fraction, we 
follow the largest cluster during the simulation run, as 
represented in Figure~\ref{fig:clusters}.
When $\phi=0.54$ (left-most panel in Figure~\ref{fig:clusters}), 
we observe a rather homogeneous and compact cluster,  whereas
more ramified structures emerge at higher  
$\phi$ (e.g. $\phi=0.56, 0.58, 0.61$). An intermediate case is 
$\phi=0.55$ (second panel from the left in Figure~\ref{fig:clusters}) 
clearly showing small compact clusters connected to each other via thin branches.

As we have already discussed elsewhere~\cite{PuseyRS,sanzPRL}, 
there is a strong correlation between 
the crystallization mechanism and the  packing fraction of the system. 
For $\phi = 0.54$, crystallization appears to follow the classical nucleation theory (CNT) picture where 
a critical cluster needs to form in order for 
crystallization to proceed.  Once a crystalline cluster overcomes a critical size, it grows in an almost  
compact structure and the entire system 
crystallizes (``classical nucleation and growth'' regime). 
Having used a large system and being at a relatively high supersaturation 
(with respect to freezing), we detect more than one critical cluster in the system (of the order of 2-3): all these clusters 
are compact, can form anywhere in the system, grow independently and, eventually, merge together.
When  $\phi \ge 0.56$ (``spinodal-like crystallization'' regime), a random  crystal growth habit is reached, where
crystallization takes place everywhere in the system. The free-energy 
barrier of nucleation gets closer to zero the higher the packing fraction, and no critical cluster size needs to be exceeded 
 in order for crystallization to proceed. 
Crystallization proceeds via 
branching of ramified clusters as $X$ grows (right-most snapshot in Figure~\ref{fig:clusters}) and 
this mechanism is reminiscent of a percolation transition associated with the crystallization process: moreover, 
at $\phi>0.58$,  crystallization happens without particles diffusing more than a diameter~\cite{sanzPRL}.
As said above, the case at $\phi=0.55$, where branching connections of more compact (and finite) clusters are found,  
is intermediate between $\phi=0.54$ and higher packing fractions ($\phi >0.56$).

\begin{figure}[h!]
\centering
\includegraphics[width=.5\textwidth,clip=]{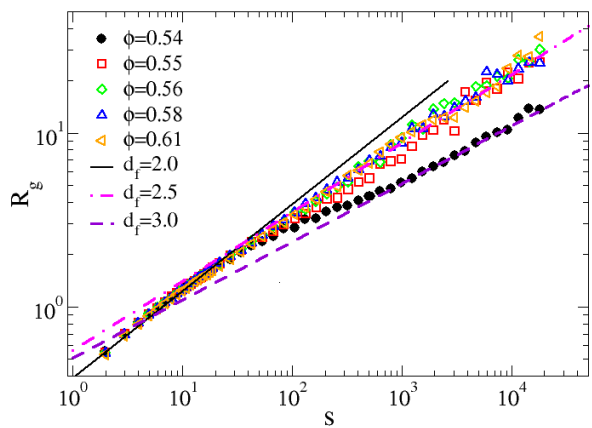}
\caption{Radius of gyration ($R_g$) versus clusters-size ($s$) at various $\phi$ ($s$ 
is averaged over different runs and over time). 
As a guide to the eye, we plot $R_g \sim  s^{1/d_f}$, with  $d_f=2.0$ (continuous black curve), 2.5 (dashed magenta curve) 
and 3.0 (dashed blue curve).}
\label{fig:radgyr}
\end{figure}

To quantify the structure of the crystalline clusters, we measure their radius 
of gyration $(R_g)$  as a function of their sizes at all packing fractions and extract 
from it an effective fractal dimension $d_f$, via $R_g \sim s^{1/d_f}$ (see Figure~\ref{fig:radgyr}).
To improve the statistics, data are averaged over all equal-sized clusters, over 
10 trajectories and over time. 

In Figure~\ref{fig:radgyr}  we observe that, at all packing fractions, the radius of gyration for small clusters follows $R_g \sim s^{1/2}$.
When clusters become larger ($s > 50$), 
their fractal dimension starts to be larger than 2 and   
their internal structure changes  depending on $\phi$.
Two distinct behaviours emerge: one pertaining to  $\phi=0.54$ 
and another to $\phi \geq 0.56$ (as shown by the superposition of the data in Figure~\ref{fig:radgyr}).
When $\phi=0.54$, crystallization follows the standard 
CNT picture, i.e. the system has to overcome a free-energy barrier in order 
to transform into a crystal. Hence, nucleation is a rare event 
and clusters appear and disappear in a stochastic way until the critical 
size is reached. The critical cluster size at this $\phi$ is estimated with CNT to be of the order of 
$s_c \sim 50$~\cite{PuseyRS}. As it can be observed from Figure~\ref{fig:radgyr}, beyond this critical value clusters show a marked change in their fractal dimension, 
becoming more and more compact  (value of $d_f$ increasing). When the cluster size exceeds $\approx 300$,  
the data clearly follow the exponent $d_f=3$, characteristic of 
a compact cluster (such as the corresponding cluster in Figure~\ref{fig:clusters})~\cite{cluster054}. 
Hence, we conclude that, at this $\phi$ (and in the CNT-regime), the structure of the clusters is mainly determined by their size.

When $\phi \geq 0.56$, nucleation is not an 
activated process any more (there is no free-energy barrier to overcome~\cite{PuseyRS})
and all $R_g$ data scale onto the same master curve:  
after a loose/open growth of the small clusters, as soon as 
$s >50$, clusters grow following the scaling predicted for random percolation in 3D
 with fractal dimension $d_f \sim 2.5$. 
This is consistent with the snapshots presented in Figure~\ref{fig:clusters}, 
where already at $\phi=0.56$ clusters start to be ramified and crystallization 
occurs with the growth of clusters appearing almost everywhere in the system. 
However, as we will discuss in Section~\ref{sec:csd}, clusters do not grow completely at random; 
instead they tend to form in the vicinity of already existing ones~\cite{sanzPRL}. This is even more evident at packing fractions 
beyond $\phi=0.58$, due to particles' lack of diffusion on the scale of their radius.

Finally  we discuss the case $\phi=0.55$, where we observe 
signatures of an intermediate behaviour. 
Given that the nucleation free-energy barrier is 
quite low, clusters start growing everywhere in the system, create a branch-like structure 
when they grow larger (as shown in the snapshot of Figure~\ref{fig:clusters}),  
and $R_g$ shows an intermediate behaviour between the compact scaling of the CNT regime and the fractal one pertaining to the spinodal-like regime (as shown in Figure~\ref{fig:radgyr}).

\subsection{Percolation analysis}
\label{sec:percolation}

We now perform an analysis of the percolation of the largest crystalline cluster in the system 
for all studied packing fractions, identifying both the percolation time $\tau_p$ and the value of the crystallinity at percolation $X_p$.
\begin{figure}[h!]
\centering
\includegraphics[width=.47\textwidth,clip=]{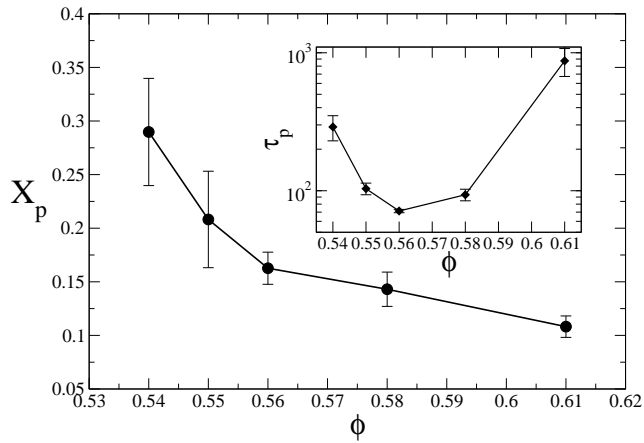}
\caption{Fraction of crystalline particles at the percolation 
threshold  $X_p$ as a function of  $\phi$.  Inset: percolation time $\tau_p$  
as a function of $\phi$.}
\label{fig:percolation1}
\end{figure}

In Figure~\ref{fig:percolation1} we report the dependence of $X_p$ on $\phi$  and observe that, for increasing values 
of $\phi$,  the crystallinity at the percolation transition has a monotonically 
decreasing  behaviour, that is steeper for $\phi <  0.56$ than for $\phi \ge 0.56$. 
To interpret this result we recall our previous findings that 
clusters are more compact with decreasing $\phi$   for $\phi  <  0.56$. Hence,  in order to percolate,  
 the largest cluster needs to reach a larger size than the one needed at higher $\phi$. On the other hand, when $\phi \ge 0.56$ 
 $d_f$ remains constant so that $X_p$ does not change much with $\phi$. 
Due to the  stochastic nature of nucleation at $\phi=0.54$, 
that  results in a small number of growing clusters that is different 
for each independent run, the $X_p$ error bar is larger at the lowest packing fractions.

For $\phi \ge 0.56$ crystalline clusters are found to percolate 
when the total crystallinity of the system is $X_p \approx 0.10-0.15$.
The fact that the overall crystallization needs to be at least 0.1 in order to observe percolation 
of the crystalline clusters (made of purely repulsive spheres) has an analog 
in studies of percolation of attractive clusters (made of particles  interacting 
via  short-range attraction and screened electrostatic repulsion)~\cite{zaccaclu}. In fact, it has been shown that 
in such systems, when the colloidal packing fraction $\phi_c$ is about 0.10, 
attractive clusters (whose size distribution obeys a power law with 
an exponent typical of random percolation) percolate throughout the system.

So far, we have presented only {\it static} observations of the structure of 
the growing clusters.
In the inset of Figure~\ref{fig:percolation1} 
we plot {\it kinetic} results of the clusters growth, representing 
the percolation time $\tau_p$ as a function of $\phi$. 
Contrary to the monotonic decrease of $X_p$, $\tau_p$ displays a clear 
minimum at intermediate $\phi$. The fastest percolation of the largest cluster is 
at $\phi=0.56$ when the largest cluster is already 
branched but particles can still easily diffuse. Whereas at low $\phi$ the percolation time is longer 
due to the compact structure of the growing cluster (requiring a larger $X_p$),  at large $\phi$ 
percolation slows down due to the slowing of particles dynamics.
This behaviour is remarkably similar to 
that observed for the ``nucleation time'' $\tau_{n}(\phi)$, defined 
as  time at which $X=0.2$, that corresponds to the time when the system 
is strongly committed to crystallize~\cite{PuseyRS}: 
at low $\phi$, nucleation is slow (large $\tau_n$) due to a high nucleation free-energy barrier, 
while at large $\phi$ nucleation slows down 
again due to the slowing particle dynamics; 
the fastest nucleation (smallest $\tau_n$) occurs around $\phi=0.56$, where the 
free-energy barrier is very low and diffusion still possible.
Therefore, the percolation of the growing clusters 
is correlated to 
the commitment of the system to fully crystallize. 
We stress that at large $\phi$, despite the slowing down of particles dynamics, 
 percolation is achieved with lower crystallinity 
than at  lower $\phi$. Hence, the kinetic slowing down is probably responsible for the continuous decrease of $X_p$, even when the shape of the clusters (quantified by $d_f$) stops changing. Indeed,   we find that  $X_p(\phi=0.61)$ is almost a factor of 3 smaller than  $X_p(\phi=0.54)$, whereas $\tau_p(\phi=0.61)$ is around 3 times larger than  $\tau_p(\phi=0.54)$, 
due to the slowing down of the dynamics beyond the  glass 
transition (occurring at $\phi \sim 0.58 $~\cite{peterbill,ZaccarelliPRL}).

\begin{figure}[h!]
\centering
\includegraphics[width=.43\textwidth,clip=]{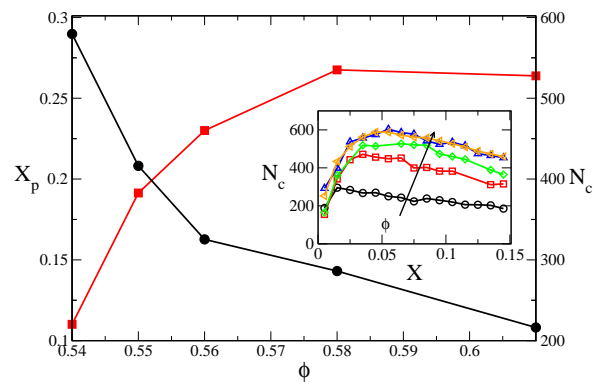}
\caption{$X_p$ (black curve) and $N_c(X=0.10)$
(red curve) as a function of $\phi$. Inset: $N_c$ as a function of $X$ (averaged within 
a $\Delta X=0.01$ interval and over independent trajectories) at different $\phi$ (from bottom to top curve: 
$\phi=$0.54, 0.55, 056, 0.58, 0.61.)}
\label{fig:numberclustersnew}
\end{figure}

The behaviour of $X_p$ with $\phi$ can also be rationalised by looking at the total number of clusters $N_c$.  In the inset of 
Figure~\ref{fig:numberclustersnew},  we report $N_c$ as a function of $X$ for different packing fractions.

It is evident that, for a given $X$, the number of clusters increases with increasing $\phi$, due to the reduced free-energy barrier, 
and it saturates when $\phi\ge 0.58$.
In particular, for $\phi=0.54$, the maximum number of clusters occurs at very small values of $X$, 
where  one or more critical nuclei have already formed 
and grow by single particles or small clusters attaching to them (that explains the slow decrease of $N_c$ with increasing $X$). 
 On the other hand, with increasing $\phi$, $N_c$ keeps increasing for small values of $X$
(since many small clusters can easily form) until it starts decreasing from  $X\sim 0.03-0.05$
(where  clusters start to branch).  We also observe   a clear dependence on $\phi$ below the glass transition ($\phi=0.58$).
This can be better seen by looking at $N_c$ for a fixed value of $X$, e.g. $X=0.10$, which is reported in 
 Figure~\ref{fig:numberclustersnew}  together with $X_p$ 
 as a function of $\phi$. Comparing the behaviour of $N_c$ to $X_p$, we can rationalise 
 the decrease of the latter as being related to the increase of the total number of clusters
up to $\phi \sim 0.58$. For larger $\phi$, $N_c$ remains constant 
and the decrease of 
$X_p$ could be partially ascribed to the increase of packing fraction, 
so that even if the number of the clusters is the same, 
the number of clusters per unit volume is higher and percolation happens at lower $X$. Moreover, minor differences 
within the statistical error of our analysis 
of the clusters shape and the cluster size distribution (see next Section) 
 could also contribute to the decrease of $X_p$ from $\phi=0.58$ to $\phi=0.61$.

\subsection{Cluster size distribution}
\label{sec:csd}


In this section we discuss the behaviour of the cluster size distribution. $n(s)$ represents 
the number of clusters of size $s$ and $f(s)$ the fraction of such clusters, i.e. $n(s)$ divided 
by the total number of clusters $N_c$. $n(s)$ varies as crystallization proceeds. To improve the statistics, we 
average over 10 independent trajectories and within $X$ intervals of width 0.01.
For all packing fractions we calculate $n(s)$ and $f(s)$  for crystallinity up to $X=0.15$; at this value, crystalline clusters 
have not percolated yet at $\phi=0.54$ and $\phi=0.55$ 
(Figure~\ref{fig:percolation1}) and have just percolated (on average) for $\phi\ge 0.56$. 
However, in the analysis that follows we have excluded percolating clusters.

\begin{figure}[h!]
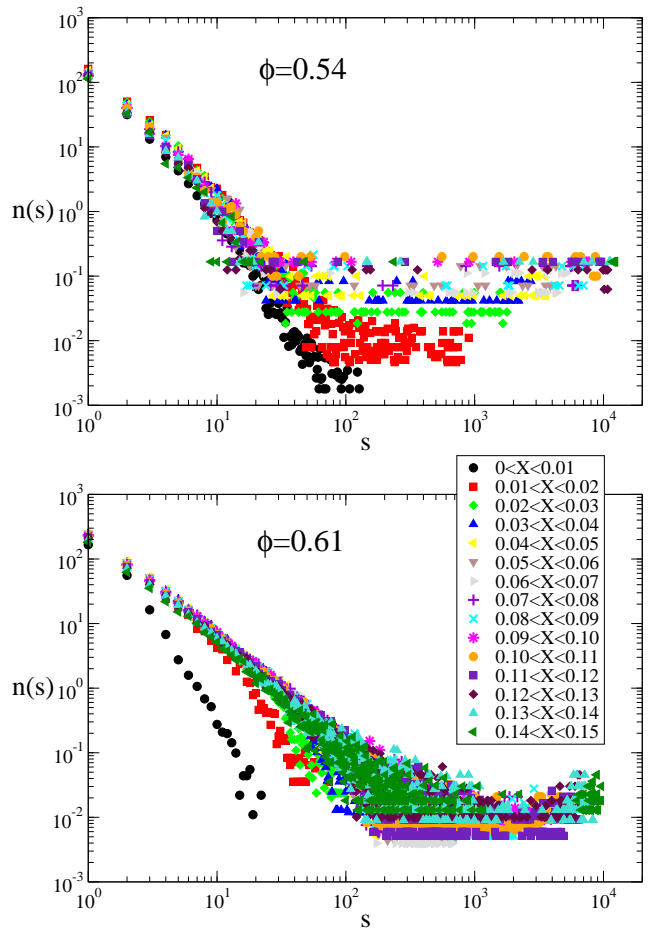

\centering
\includegraphics[width=.47\textwidth,clip=]{lin-plot-054-2}
\includegraphics[width=.47\textwidth,clip=]{lin-plot-061-2}
\caption{Number of clusters $n(s)$
at $\phi=0.54$ (top) and $\phi=0.61$ (bottom).
Each data set is averaged over 
10 independent runs in intervals ranging from $0.0 \le X \le 0.01$ to $0.14 \le X \le 0.15$ ($\Delta X=0.01$). $n(s)$ can take values smaller than 1, being computed as the number of clusters for each value of $s$ divided by
the number of runs and by the number of configurations analysed for a given 
$X$ window. 
Symbols are the same for both panels. 
Only non-percolating clusters are considered in the analysis.\label{fig:ns5461}}
\end{figure}

In Figure~\ref{fig:ns5461} we report  $n (s)$ 
 for different $X$-intervals both for $\phi = 0.54$ (top) and $\phi = 0.61$ (bottom). 
For both packing fractions,  $n(s)$ shows a rapid decay for $X<0.01$
signaling the growth of the first few crystalline clusters. 
Once $X\ge 0.03$ (when $\phi=0.54$) and $X\ge 0.06$ (when $\phi=0.61$),
all curves collapse onto each other,  showing that the clusters distribution has reached a stationary
 profile during crystal growth. This happens for all studied $\phi$ and justifies further averaging of the $n(s)$ curves over all configurations within the range $0.06 < X < 0.14$~\cite{averagingX}.
We also notice that large clusters, e.g. $s\gtrsim 100$, are rare in the case of $\phi=0.54$ while their number becomes very large for $\phi=0.61$, as indicated respectively by scarce/abundant population of large-$s$ points in Figure.~\ref{fig:ns5461}. 

\begin{figure}[h!]
\centering
\includegraphics[width=.47\textwidth,clip=]{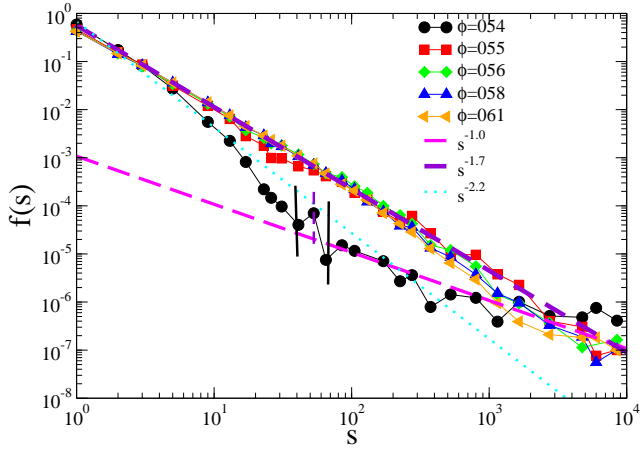}
\caption{Fraction of clusters of size $s$, $f(s)$, averaged over the $X$-intervals 
where the curves collapse ($0.06 < X \le 0.14$ at all $\phi$). 
As a guide to the eye, we also plot the power-law dependence of $s^{-1.0}$ (dashed magenta curve), $s^{-1.7}$ (dashed violet curve)
and $s^{-2.2}$ (dotted cyan curve). 
Only non-percolating clusters are considered in the analysis.  The vertical lines $s^*=50 \pm 10$  indicates the crossover between two regimes at $\phi=0.54$
\label{nsav}.}
\end{figure}

We now focus on the  fraction  of clusters of size $s$, $f(s)$.
In order to evaluate $f(s)$ for increasingly large $s$ (where clusters of each size become more and more scarce),  
we apply the following criterion: 
(i) we arbitrarily subdivide the range of $s$-values  into 
intervals $\Delta s_i=(s_{i,max}-s_{i,min})$ containing at least one cluster each and estimate 
$n^t_i$, i.e. the total number of clusters within each interval $\Delta s_i$; (ii) 
we assign the value of $n_i=n^t_i/\Delta s_i$ to every $s$ within $\Delta s_i$. The total 
number of clusters can be computed as $N_c=\sum_i n_i  \Delta s_i$ and the fraction of clusters 
of size $s$, $f(s) = n_i/N_c$. We now represent $f(s)$ as a function of $s$, 
the closest integer to the central value of $\Delta s_i$.
We leave to the Appendix~\ref{sec:deltas} the definition of all  $\Delta s_i=(s_{i,max}-s_{i,min})$ intervals.
In Figure~\ref{nsav} we plot $f(s)$ for several packing fractions ranging from 0.54 to 0.61.
Data are averaged for $0.06 < X < 0.14$ (within this range we have found the profiles to be stationary and independent of $X$~\cite{averagingX}). 

First of all, we notice the marked difference that is found between the data corresponding to $\phi=0.54$ and all other data. This demonstrates that the stationary 
cluster size distribution is, within our statistical uncertainty, identical for any $\phi$ 
larger than $0.55$.
This behaviour is similar to what we 
observed for $R_g$, pointing to the universality of (static)  properties of the clusters once 
spinodal-like nucleation has occurred.

From the calculated $f(s)$ we can search for the emergence of a power-law behaviour approaching crystallization, in analogy with standard attractive systems approaching a percolation transition. We find that when $\phi = 0.54$ 
a crossover between two regimes takes place at $s^* =50$, 
which corresponds to the estimate of the critical cluster size according to CNT~\cite{PuseyRS}.  
For $s < s^*$  the cluster distribution is not too different from that of a system approaching random percolation ($f(s) \sim s^{-2.2}$); 
whereas  larger clusters grow in a  compact way, resulting in a much lower power-law exponent ($f(s) \sim s^{-1.0}$). 
When $\phi \ge 0.55$, $f(s)$ seems to follow a 
power law with exponent $\tau \sim 1.7$ for the entire size range (even for very small cluster sizes). 
We notice that this value for $\tau$ is similar to that of the exponent found for the size distribution of mobile regions in glass-forming systems~\cite{glotzer}.
The discrepancy with the exponent expected from random 
percolation theory ($\tau \sim 2.2$) may stem from the fact that  
new crystalline regions preferably appear in the surroundings of existing ones~\cite{sanzPRL}, 
which causes a partial loss of randomness of the growing  aggregate.



\begin{figure}[h!]
\centering
\includegraphics[width=.47\textwidth,clip=]{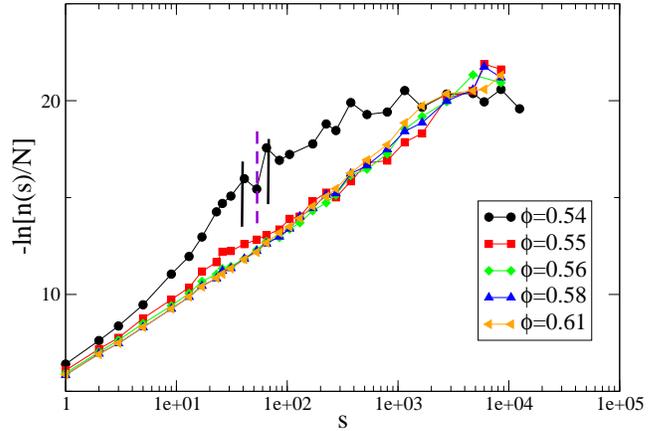}
\caption{$-\ln(n(s)/N)$ versus $s$ for $0.54 \le \phi \le 0.61$. 
$n(s)/N=f(s)\times N_c/N$ (where $f(s)$ data are  from Figure~\ref{nsav}). 
 The vertical lines $s^*=50 \pm 10$  indicates the crossover between two regimes at $\phi=0.54$ \label{lnnN}.
}
\end{figure}

We now analyse further the behaviour of $f(s)$ for $\phi = 0.54$ by using CNT, where a clear crossover between two regimes
is present. When $s$ is small,  sub-critical clusters form and re-dissolve 
whereas when $s$ is large  post-critical clusters irreversibly grow.  Only 
the large-$s$ regime seems to follow convincingly a power law with exponent $\tau \sim 1$, whereas
the small-$s$ regime starts deviating from random aggregation ($\tau \sim 2.2$) when $s$ approaches $s^*$.
According to CNT~\cite{kelton},  the free-energy barrier associated to the appearance of
 size-$s$ clusters is written as 
$\beta \Delta G(s)=  - \ln \left( P(s) \right)$, 
where $P(s)= n(s)/N$ is the probability to have size-$s$ clusters in the system.
Therefore, we can  use  the cluster size distribution $n(s)/N$ (or equivalently $f(s) \times N_c/N$)
to estimate the barrier, as shown in Figure~\ref{lnnN} for different packing fractions
(averaged over the range $0.06 < X < 0.14$).

At all packing fractions higher than $0.55$ the curves in Figure~\ref{lnnN} collapse, showing  that crystallites 
starts growing in the same way whenever $\phi>0.55$:  
multiple crystalline patches appear 
and crystallization proceeds by a sequence of stochastic micronucleation events, correlated in space by emergent dynamic heterogeneity~\cite{sanzPRL}.
As we have already mentioned, the case when  $\phi =0.55$ is intermediate between 
$\phi>0.55$ and $\phi=0.54$. When $\phi=0.55$ we cannot distinguish between two regimes and 
we interpret this as a consequence of the practical absence of a barrier-crossing nucleation event. 
Whereas when $\phi = 0.54$  we observe  two regimes: one characterised by pre-critical clusters  ($s< s^*$) and 
 one by post-critical clusters at large $s$. 
Therefore, in the small $s$ regime, $-\ln(n(s)/N)$ can be interpreted as the free energy of cluster
formation 
 whose top can be read from the continuous vertical lines in Figures 6 and 7, providing an estimate for $\Delta G \sim 16$.
This  value is consistent with the value of the top of the free energy barrier 
 calculated in Ref.~\cite{auer,filion1} for a lower packing fraction ($\phi \sim 0.535$):  $\beta \Delta G^* \approx 20$.
This consistency check makes us gain confidence of our cluster size distribuition analysis.

Within Classical Nucleation Theory, once we know the top of the free energy 
barrier $\beta \Delta G^*$ and the size of the critical nucleus $s^*$, 
it is possible to estimate  the 
difference in chemical potential between the fluid and the solid:
$\beta \Delta \mu_{CNT} = \frac{2 \beta \Delta G^*} {s^*} $.
The value of $\Delta \mu_{CNT}$ is order-parameter dependent given that, 
contrary to $\beta \Delta G^*$, $s^*$ depends on the choice of the order parameter 
used to identify solid-like particles~\cite{filion1}.
Only for particular choices of the order parameter does $\Delta \mu_{CNT}$ coincide 
with the  true value of $\Delta \mu$ (obtained, for instance, from thermodynamic
integration). 
In our case, 
if we use  the CNT expression for $\beta \Delta \mu_{CNT}$ taking the value of the top of the free-energy barrier and the critical nucleus 
size from Figure~\ref{lnnN}($\beta \Delta G^*= 16$ and $s^*=50$), we obtain
$\beta \Delta \mu_{CNT} = 0.64$, which is in good agreement with
the value  computed 
from thermodynamic integration $\beta \Delta \mu=0.63$ in Ref.~\cite{filion1}.
Thus, our choice of the order parameter to identify 
crystalline particles seems to be reasonable.

\subsection{Effect on crystallization of the cutoff needed to define of a solid particle}
\label{sec:nsc}

\begin{figure}[h!]
\centering
\includegraphics[width=.37\textwidth,clip=]{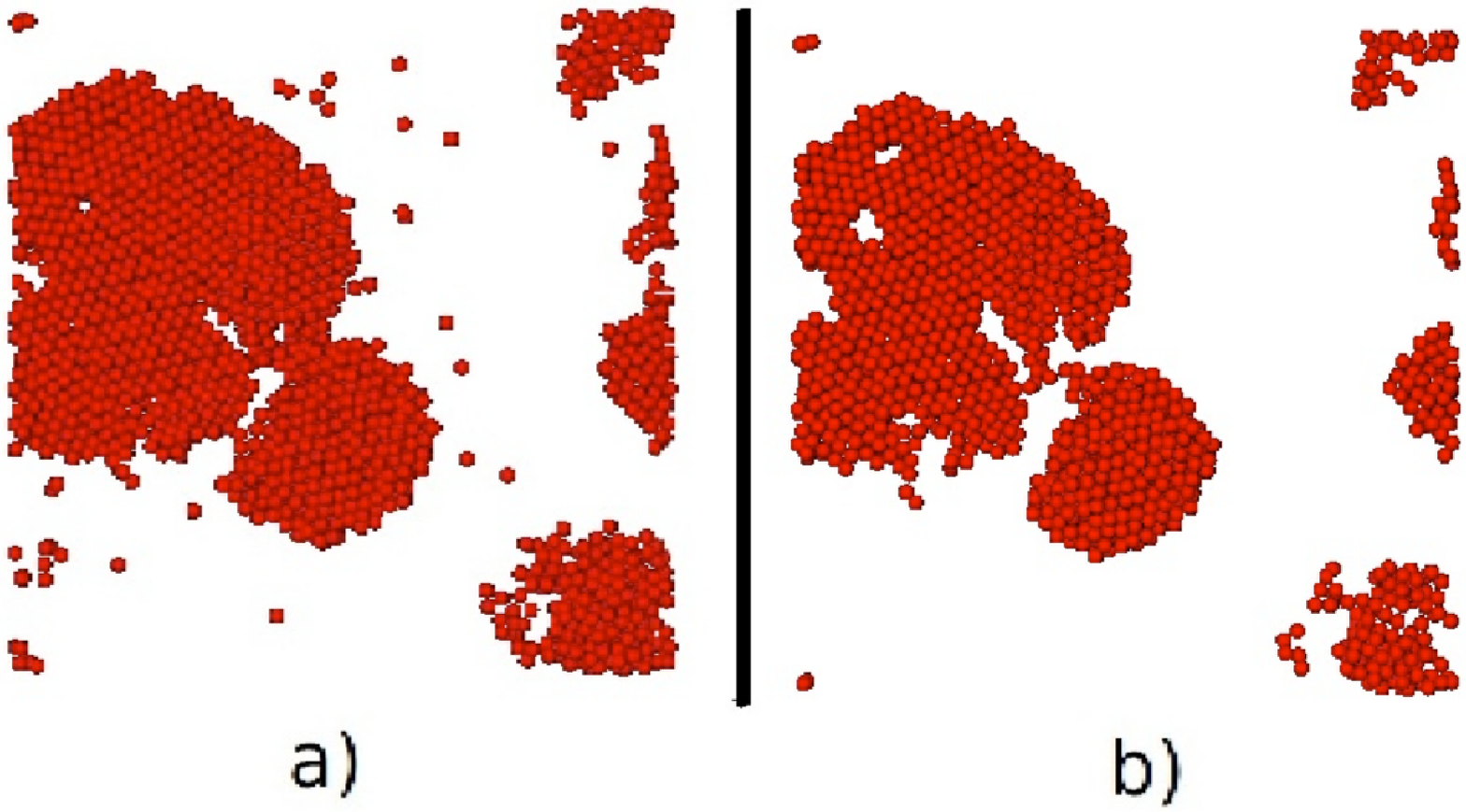}\\
\includegraphics[width=.37\textwidth,clip=]{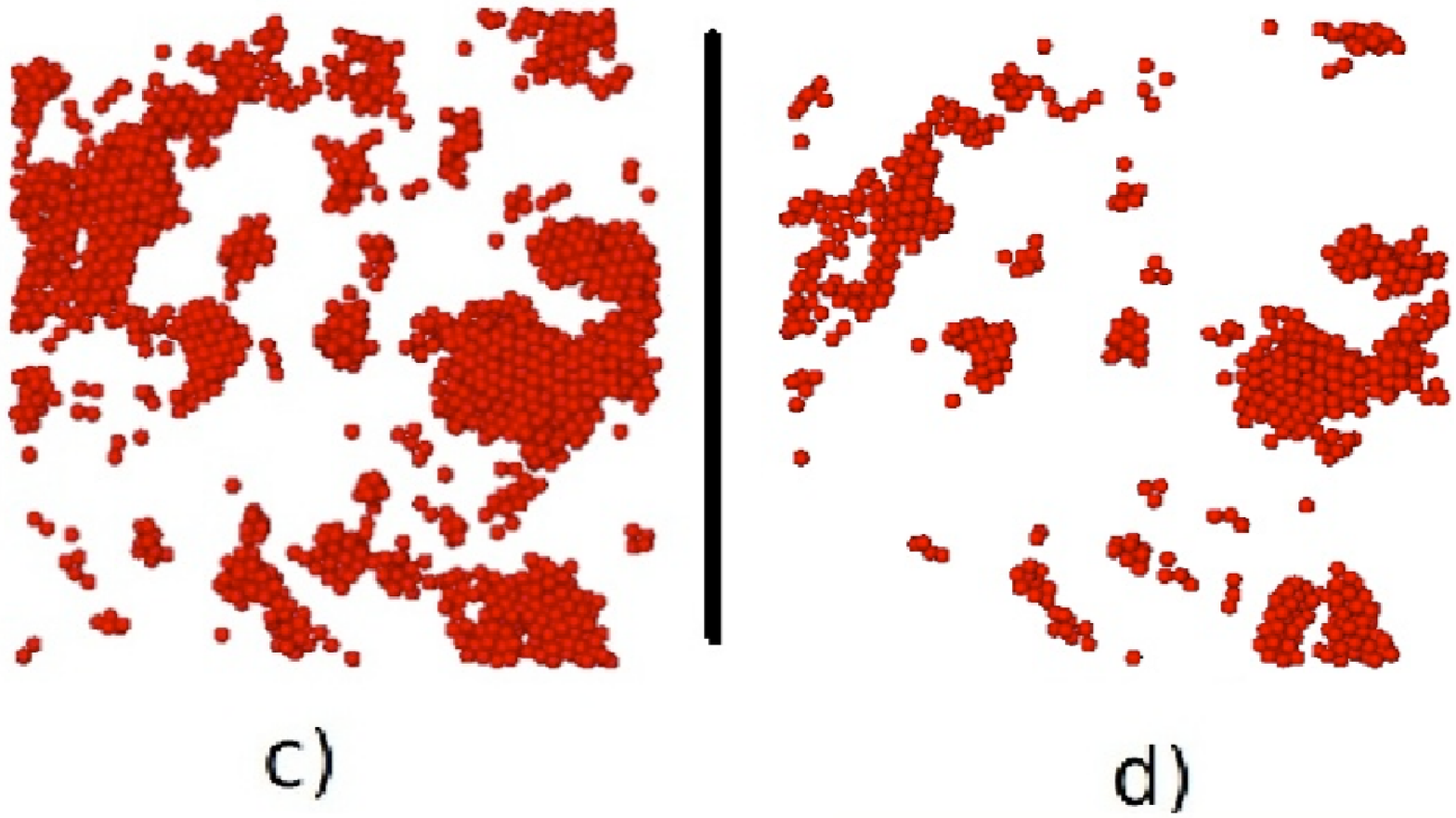}\\
\caption{Slabs (about 3-4 particles diameters thick) 
where solid particles are identified with: Top: $\phi=0.54$, a) $\xi_c=6$ ($X\sim 0.20$) and b) $\xi_c=10$ ($X\sim 0.10$); 
Bottom: $\phi=0.61$, 
c) $\xi_c=6$ ($X\sim 0.20$) and d) $\xi_c=10$ ($X\sim 0.10$).}
\label{fig:clustNsn}
\end{figure}

We now study the effect of the choice of $\xi_c$ (the parameter used to define solid-like
 particles in our definition of crystalline clusters),  
on  the properties of the clusters, such as the 
radius of gyration.
We focus on the two extreme cases  studied, $\phi=0.54$ and $\phi=0.61$.
To identify the neighbours of each particle, we use the criterion of Ref.~\cite{koos} 
and determine the number of connections with $d_6>0.7$. 
At this point, we use various definitions of the 
number of connections needed for a particle to be identified as solid-like: 
not only $\xi_c=6$ (what we have used so far to identify a solid-like particle), but also 
$8$ or $10$.
To qualitatively understand the effect of the choice of $\xi_c$ on the structure 
of the growing clusters, we represent 
a slab of the system taken at the same position in the sample   
when  $\phi=0.54$ (Figure~\ref{fig:clustNsn} a and b) 
and  when  $\phi=0.61$ 
(Figure~\ref{fig:clustNsn} c and d). 
In panels a) and c) solid particles are evaluated 
using $\xi_c=6$ and correspond to a system with $X\sim 0.20$
whereas in panels b) and d) they are calculated using  $\xi_c=10$ and 
correspond to a system with $X\sim 0.10$.

From Figure~\ref{fig:clustNsn}, it is clear that the overall structure of the clusters 
is not much affected by the choice of $\xi_c$. 
Filion et al.~\cite{filion1} have recently evaluated the effect of the
choice of $\xi_c$ on the calculation of the top of the nucleation free-energy barrier $\Delta G^*$ at
$\phi=0.535$ (see Figure 2 in Ref.~\cite{filion1}), showing that $\Delta G^*$ does not
depend on $\xi_c$ (named in the same way in their work). This contrasts with  the corresponding value
of the size of the critical cluster $s^*$: for $\xi_c=6$ they find $s^* \sim 100$, whereas for
$\xi_c=10$ they obtain $s^* \sim 25$.
The authors conclude that the main difference
among the order parameters (each defined with a given value of $\xi_c$, $5 \le \xi_c \le 10$)
is the ability to  distinguish between fluid--like and solid--like particles near the fluid--solid interface.

\begin{figure}[h!]
\centering
\includegraphics[width=.47\textwidth,clip=]{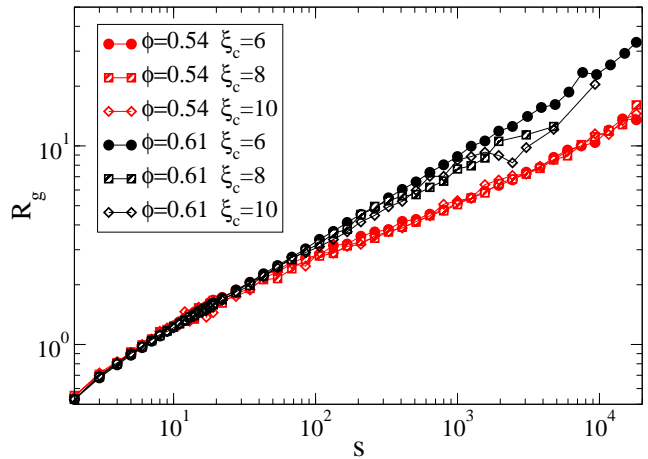}
\caption{$R_g$ versus $s$ at $\phi=0.54$
(red) and $\phi=0.61$ (black) averaged over time and over 10 runs. In both cases,  filled circles, 
striped squares and empty diamonds correspond to clusters where solid-particles are 
defined using $\xi_c=6$, $\xi_c=8$, and $\xi_c=10$, respectively.
The spreading of the data for large values of $s$ at 
$\phi=0.61$ is due to poor statistics for 
very large clusters. }
\label{fig:radgyrmanycutoff}
\end{figure}

To quantify our observations, we 
represent the radius of gyration as a function of the cluster size 
at packing fraction $\phi=0.54$ and $\phi=0.61$ (Figure~\ref{fig:radgyrmanycutoff}) for different values of $\xi_c$.

Results for $\phi=0.54$ and $\phi=0.61$ clearly 
provide evidence that $R_g$ does not depend strongly on the different 
values of $\xi_c$ used in the definition of solid-like particles.
The slabs of Figure~\ref{fig:clustNsn} corroborate the result presented 
in Figure~\ref{fig:radgyrmanycutoff}, and we conclude 
that the structure of the  growing crystalline clusters is not affected by the choice of 
 $\xi_c$.




We now re-plot Figure~\ref{fig:clustNsn} assigning different colours to each particle depending whether
a particle has been labelled as solid-like with one of the above-mentioned criteria:
in red, we colour particles that are solid-like according to the most stringent criterion ($\xi_c=10$),
in orange particles identified as solid-like using $\xi_c=8$ (but not with $\xi_c=10$) 
 and in yellow particles  identified as solid-like using $\xi_c=6$ (but not with $\xi_c=8$).
\begin{figure}[h!]
\centering
\includegraphics[width=.47\textwidth,clip=]{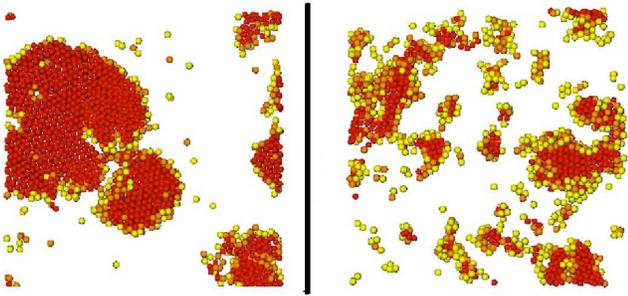}\\
\caption{Left: $\phi=0.54$. Re-plot of Figure~\ref{fig:clustNsn}a), where we colour  in red  particles
identified as solid-like with $\xi_c=10$ (corresponding to the ones represented in Figure~\ref{fig:clustNsn}b),
in orange particles identified as solid-like using $\xi_c=8$ (but not with $\xi_c=10$) and
in yellow particles identified as solid-like using $\xi_c=6$ (but not with $\xi_c=8$).
Right: $\phi=0.61$. Re-plot of Figure~\ref{fig:clustNsn}c) with the same colour-code. 
Particles identified as solid-like with $\xi_c=10$ correspond to the ones represented in Figure~\ref{fig:clustNsn}d). The colours become black, dark grey and light grey, respectively, in black-and-white.\label{fig:tancol}}
\end{figure}

When $\phi=0.54$ (Figure~\ref{fig:tancol}(left)) we observe that the criterion with $\xi_c=10$
allows one to obtain particles in the inner part of the clusters,
surrounded by particles defined as solid-like according to the looser criterion
$\xi_c=8$ and then by the ones defined with $\xi_c=6$.
During the time-evolution of the clusters, we always observe the colour-code
ranging from red (black) in the inside to yellow (light grey)  in the outside of the clusters.
As we have already discussed, small clusters for $\phi=0.54$ as well as branched ones
for $\phi=0.61$ are far from being compact: in these cases, most particles will live at the interfaces, so that red (black) particles are less abundant.

In recent works~\cite{schope,schilling,john}, it has been suggested that 
in hard sphere systems at low packing fractions ($\phi\sim 0.54$) 
nucleation happens 
via a two-step scenario: first the formation of a locally denser regions with high bond orientational order 
parameter, and next the restructuring of these regions at constant density.
Analogously, either at $\phi=0.54$ and at $\phi=0.61$, we first observe 
the formation of loosely packed clusters (in yellow), that become
more compact when growing (in orange) and within which
we identify the growth of highly ordered regions (in red). 
However, from our analysis 
one could argue that nucleation happens via 
a ``multi-step'' mechanism, and that the number of steps depends on the number of cutoffs $\xi_c$ that are monitored, i.e. three in this case.
In our previous work~\cite{sanzPRL}, we have demonstrated how crystallization
from a glass (at $\phi=0.61$) happens via a gradual (many-steps) mechanism, but we did
not discuss the process at packing fractions below the glass transition
($\phi_g=0.58$). As we have just shown, the mechanism of gradual growth
seems to hold also in super-saturated systems just above freezing.

\subsection{Discussion and conclusions}

In this paper we have analysed the properties of clusters of solid particles 
during crystal growth in a system of monodisperse hard spheres. We have focused on several 
packing fractions beyond the fluid-crystal transition up to state points deeply in the glassy regime.
Given the marked differences in the nucleation processes taking place at different $\phi$, i.e. from 
CNT regime up to crystallization from a glass~\cite{PuseyRS,sanzPRL}, we expected 
to observe clear differences also in the properties of the growing clusters.
 On the basis of the results presented here, we can clearly identify two regimes where clusters are characterised
 by distinct statistical properties: one pertains to $\phi \le 0.54$ and one to $\phi \ge 0.56$. The value $\phi=0.55$
  marks the threshold between the two behaviours, showing intermediate properties.

In the low-$\phi$ regime ($\phi=0.54$), CNT holds and the existence of a critical size $s^*$ 
leads to a cluster size distribution which shows two distinct behaviours respectively below and above $s^*$ (at long enough times). 
The structure of the clusters is loose for very small sizes ($s < 50$) and crosses to spherical and 
compact ($d_f=3$) for larger sizes (this can also be visualised in the snapshots shown in Figure~\ref{fig:clusters}).
When studying the cluster size distribution, 
we find analogies between crystalline clusters growing  and 
  standard clusters of attractive particles approaching percolation: 
the cluster size distribution seems to approach a power-law behaviour also in our case, albeit with markedly different exponents. 
At $\phi=0.54$ and for $s < s^*$, the cluster distribution
it is not too different from a random-percolation power-law, i.e. $f(s) \sim s^{-2.2}$. 
However, for $s > s^*$,  $f(s)$ crosses to a clearer power-law behaviour, with a
 much lower effective power-law exponent, i.e. $f(s) \sim s^{-1.0}$, due to the fact that  
   larger clusters grow in a  compact way.
The use of CNT
 has allowed us to provide estimates of the free-energy barrier at $\phi=0.54$ and 
 of the difference in chemical potential between solid and liquid phase that, being in agreement with thermodynamic 
 integration calculations for the same system at the same conditions~\cite{auer,filion1}, corroborates 
 our choice of the order parameter used to identify solid-like particles.

In the large-$\phi$ regime ($\phi \ge 0.56$), the free-energy barrier to nucleation becomes negligible, so that no 
critical cluster size exists: small size clusters  have a loose structure and 
 do not reach a compact form when growing ($d_f \sim 2.5$). 
When studying the cluster size distribution, 
we observe a power-law behaviour for all sizes, but the
 exponent ($\tau\sim 1.7$) is smaller than the value predicted by random percolation  ($\tau\sim 2.2$). This could be ascribed to the fact 
that clusters do not grow completely at random, but preferably
particles become solid-like in the vicinity of solid regions, which act as seeds for further growth~\cite{sanzPRL}. 
This mechanism should become more and more important with increasing $\phi$, due to 
the reduced diffusion of the particles. However, the surprising result is that the same behaviour 
is observed for any packing fractions above $\phi=0.56$, covering a range where the dynamics 
slows down by several orders of magnitude~\cite{ZaccarelliPRL}.  Hence, we conclude that the static properties of these 
clusters are not much affected by dynamics.

We also notice a striking similarity between  percolation events and nucleation.
When $\phi<0.56$, clusters are more compact and take a long time to percolate: similarly, the nucleation time (defined as the time 
when $X=0.20$ in Ref.~\cite{PuseyRS}) is large at $\phi=0.54$, given that crystallization is an activate process, and 
decreases when $\phi=0.55$, due to the lowering of the free-energy barrier. 
When $\phi \ge 0.56$,  clusters percolate throughout the system and the nucleation time increases with $\phi$ 
due to the slowing down of particles dynamics.
Of course, the resulting macroscopic crystals at all packing fractions 
are quite different in the two nucleation regimes, as evident from the snapshots presented in Figure~\ref{fig:clusters}. 
On the one hand, 
homogeneous and compact 
structures are found in the CNT region, where crystallization once activated proceeds to the formation of a full crystal. On the other 
hand, for larger $\phi$, the macroscopic crystals that are formed are more heterogeneous, since they result from the 
branching of many sub-units. In this case, it is most likely that the final structure will have a
poly-crystalline character (with different units possibly bearing a different orientation).

We stress that, for both regimes (low and high $\phi$), the cluster properties are very different from those observed in a standard aggregating system, 
driven by attractive interactions. The main difference can be observed for small sizes, where attraction drives the growth of 
compact small clusters ($d_f=3$) \cite{mossa}.
Then, with increasing size, the structure of the clusters remains spherical upon growth if particles are interacting via a pure attraction, 
while a fractal structure (decreasing $d_f$) is observed when additional interactions, like for example long-range electrostatic repulsion, 
are also at work~\cite{mossa,zaccaclu}.
In the present case of crystalline clusters, the loose structure for the small clusters can be 
understood by the absence of a real attraction between the particles.
Hence larger clusters become compact in the classical nucleation regime (at small supersaturations),
 while they turn into random fractal objects when nucleation happens without activation (at large supersaturations $\phi \ge0.56$), so 
 that adjacent solid particles merge and branch into a percolating cluster.

%



In conclusion, we have provided new insights into the growth of clusters of solid particles during the nucleation process of monodisperse hard spheres. Our results shed more light on the existence of different nucleation regimes  already discussed in 
Ref.~\cite{ZaccarelliPRL,PuseyRS,chantaljpcm,sanzPRL}, highlighting the influence of these onto the connective  properties of the system. In particular we identify that while compact crystals are formed at small supersaturations, a crossover to the formation of fractal ramified clusters is encountered upon increasing $\phi$. 
Surprisingly, the static properties of the growing clusters in the high-$\phi$ regime ($\phi \ge 0.56$) 
are $\phi$-independent.
In our previous work~\cite{sanzPRL}, we have demonstrated how crystallization
from a glass (at $\phi=0.61$) happens via a gradual (many-steps) mechanism, but we did
not discuss the process at packing fractions below the glass transition
($\phi_g=0.58$). 
In this paper we show how the
 mechanism of gradual growth seems to hold also in super-saturated systems just above freezing. 

The two growth regimes (at $\phi<0.56$ and $\phi \ge 0.56$) may explain the different morphology of the crystals 
observed at low and high $\phi$ in the experiments of Pusey and van Megen~\cite{peterbill}.
However, to allow for a more detailed comparison 
with those experimental results, a similar investigation for slightly polydisperse hard spheres, which should 
also favour fractionation and
hence a stronger tendency to the formation of poly-crystals, will be the subject of a future investigation.

\section{Acknowledgements}
We thank C.~De Michele for the code used to generate the snapshots in Figure~1.
CV and ES acknowledge financial support from an Intra-European Marie Curie Fellowship (in Edinburgh) 
and from a Juan de La Cierva and Ramon y Cajal Fellowship, respectively (in Madrid). 
WCKP, MEC and EZ acknowledge support from ITN-234810-COMPLOIDS and WCKP and MEC the EPSRC grant EP/EO30173. MEC holds a Royal Society Research Professorship.
This work has made use of the resources provided by the Edinburgh Compute and Data Facility (ECDF). The ECDF is partially supported by the eDIKT initiative.

\appendix
\section{Local bond-order parameter \label{sec:appa}}

We use the rotationally invariant local bond order parameter 
$d_6$ defined as the scalar product between particle $i$'s $q_6$ 
complex vector and the one of each of its neighbour $j$, 
$d_6(i,j)=\sum_{m=-6}^{6} q_{6,m}(i)\cdot q^*_{6,m}(j)$ 
 (where $q^*_{6,m}$ is the 
complex conjugate), then averaged over all 
particle $i$'s neighbours, $N_b(i)$~\cite{stein,frenkel,tenwolde}.
Each component $q_{6,m}(i)$ depends on the relative orientation of 
particle $i$ with respect to its $N_b(i)$ neighbouring particles, and is 
defined as $q_{6,m}(i)=\frac{1}{N_b(i)} \left[ \sum_{j=1}^{N_b(i)} \psi_{6,m}(\theta_{i,j},\phi_{i,j})  \right] / \left[  \sum_{m=-6}^{6} |  q_{6,m}(i) |^2 \right] $ (with $m=[-6,6]$), where $\psi_{6,m}(\theta,\phi)$ are the spherical harmonics of order 6.

In order to identify particle $i$'s neighbours we use 
the criterion of Ref.~\cite{koos}, that is capable of avoiding a  density dependence cutoff distance $r_b$.
We have checked that such criterion is consistent with a cutoff distance criterion, 
where $r_b$ is tuned  for any packing fraction.
We then consider particles $i$ and $j$ as having a "solid connection" 
when their 
$d_6(i,j)$  exceeds the value of 0.7: particle $i$ is then 
labelled as solid-like if it has at least $\xi_c$=6 solid connections.

$d_6(i,j)$ is  a normalised quantity correlating the local environments of 
neighbouring particles, it is a real number and is defined in the 
range $-1 \le d_6(i,j) \le 1$:  it decreases when thermal vibrations are 
present but, on average, it is close to one if particles have a solid-like 
environment, and around zero if particles have a fluid-like environment.

\section{Values of $\Delta s_i$ used to compute  $f(s)$\label{sec:deltas}}

In section~\ref{sec:csd} we compute the fraction of clusters of size $s$, $f(s)$: 
each value of $f(s)$ is computed within a given interval of width $\Delta s_i=(s_{i,max}-s_{i,min})$. 
In Table~\ref{intervals} we explicitly indicate all $\Delta s_i$ 
used to calculate $f(s)$. In our calculations we have also checked that different choices 
of interval widths yield the same cluster size distributions, within statistical error.
\newpage
\begin{table}[h!]
\begin{center}
\caption{Bonudaries of the cluster size intervals, $[s_{i,max},s_{i,min}]$, centred at $s$.}
\label{intervals}
\begin{tabular}{lcc|cc|cc}\hline\hline                         &
$[s_{i,max},s_{i,min}]$ & $s$ & $[s_{i,max},s_{i,min}]$ & $s$ & $[s_{i,max},s_{i,min}]$ & $s$      \\ \hline
$\mbox{}$ & 1& $ 1        $ & [36,45]&$41$     & [451,600]& $525$ \\
$\mbox{}$ & 2& $ 2        $ & [46,55]&$53$     & [601,1000]&$800$ \\
$\mbox{}$ & 3& $ 3        $ & [56,75]   &$65$  & [1001,1300]&$1150$ \\
$\mbox{}$ & [4,7]& $ 5    $ & [76,95]  &$85$   & [1301,2000]&$1650$ \\
$\mbox{}$ & [8,11]& $ 9   $ & [96,120] &$105$  & [2001,3500]&$2750$ \\
$\mbox{}$ & [12,15]& $ 13 $ & [121,140]&$130$  & [3501,5000] &$4650$ \\
$\mbox{}$ & [16,21]& $ 17 $ & [141,200] &$170$ & [5001,7000] &$6000$ \\
$\mbox{}$ & [22,25]& $ 23 $ & [201,250] &$225$ & [7001,10000] &$8500$ \\
$\mbox{}$ & [26,27]& $ 26 $ & [251,300] &$275$ & [10001,15000] &$12500$ \\
$\mbox{}$ & [28,35] & $ 31$&  [301,450] &$375$ & & \\
\hline\hline
\end{tabular}
\end{center}
\end{table}


\begin{thebibliography}{99}
\bibitem{cluster054}
To make sure that the average procedure does not cause any loss of information, 
we have also plotted $R_g$ raw data (without any average) and compare them with 
the curve obtained by averaging (data not shown).
We then confirm that the average does not alter the information, and 
no modification in the cluster-structure takes place 
while averaging over time and over independent runs.
\bibitem{averagingX}
We have proved that the distributions averaged over another $X$-interval, namely $0.2 < X < 0.6$,
are identical within statistical error to the ones reported here and averaged over $0.6 < X < 0.14$. 
This reinforces the idea that during the growth of $X$ a stationary cluster size distribution emerges.
\end{thebibliography}

\end{document}